\documentclass[twocolumn,showpacs,prl]{revtex4}
\usepackage{graphicx}
\usepackage{epsfig,color}
\usepackage{amsmath}
\usepackage{amssymb}
\usepackage{gensymb}
\usepackage{bm}

\newcommand{\beq}{\begin{eqnarray}}
\newcommand{\eeq}{\end{eqnarray}}

\begin{document}

\title{Ferroelectricity induced by cooperative orbital ordering and Peierls instability.}

\author{Paolo Barone and Silvia Picozzi}

\affiliation{
Consiglio Nazionale delle Ricerche, Istituto Superconduttori,
materiali innovativi e dispositivi (CNR-SPIN), 67100 L'Aquila, Italy}

\date{\today}

\begin{abstract}
A general mechanism by which orbital ordering, coupled to Peierls-like lattice distortions,
can induce an electronic switchable
polarization is discussed within a model Hamiltonian approach in the
context of the modern theory of polarization. By using a Berry-phase approach, a clear picture emerges in terms of
Wannier-function centers and orbital occupancies. The proposed mechanism may apply to oxide spinels whose electronic
structure has effective one-dimensional character, such as CdV$_2$O$_4$, recently proposed to display multiferroic
behaviour.
\end{abstract}


\maketitle
{\it Introduction.} Since the revival of interest in the magnetoelectric effect and in the multiferroic state of
condensed matter\cite{fiebig,tokura2010,khomskii2009}, the possible electronic, as opposed to purely ionic, origin of ferroelectric polarization has
attracted, and still attracts, increasing attention\cite{batista2004,ishihara2010}.
A significant contribution to the field came from the
rigorous definition of macroscopic bulk polarization $\bm P$ in crystal lattices, as given in the early 90's, which eventually
led to the formulation of what is nowadays known as the modern theory of polarization\cite{vanderbilt93,resta94,berryp_rev}.
According to this theory, polarization
changes can be formally expressed in terms of the Berry phase (BP) of the electronic wavefunctions, which correctly captures
the contribution to $\bm P$ coming from mobile electrons in crystals with covalent character - what is generally
referred to as the electronic contribution to polarization. 
Several mechanisms responsible for electronic ferroelectricity
has been devised since,
involving spin\cite{knb,cheong2007}, charge\cite{efremov2005,khomskiibrink2008} and orbital\cite{keimer2006,barone_prl}
degrees of freedom of electrons. 

Due to its potential straight application in magnetoelectric
devices, the first class of mechanisms, that applies directly to magnetically induced ferroelectrics, has been object of intense
research\cite{tokura2010,knb,cheong2007}. Nonetheless, our understanding of general mechanisms leading to electronically
induced ferroelectricity cannot neglect the implications of charge- (CO) or orbital-ordering (OO) phenomena. The essential mechanism
by which a charge ordering can lead to a bulk polarization is fairly well understood
\cite{khomskiibrink2008,khomskii2009}. Ferroelectricity, in fact, may arise due to a simultaneous inequivalency of both
sites and bonds after charge ordering, a situation often found in transition metal compounds containing transition metal
ions with different valence. Bond inequivalency may be structural, as for example in quasi one-dimensional organic
materials (such as (TMTTF)$_2$X\cite{tmttf} or TTF-CA\cite{gianluca-ttfca}), or may have a magnetic origin, as in the
case of Ca$_3$CoMnO$_6$\cite{choi} or $R$NiO$_3$\cite{gianluca-nick}, where ferromagnetic and antiferromagnetic bonds
differentiate through symmetric exchange striction.
The simultaneous presence of site-centered and bond-centered CO has been also proposed to induce ferroelectricity in half-doped
manganites such as Pr$_{0.5}$Ca$_{0.5}$MnO$_3$ or La$_{0.5}$Ca$_{0.5}$MnO$_3$\cite{efremov2005}. In this case, however, we
have suggested that orbital degrees of freedom may play a relevant role in establishing the ferroelectric state\cite{barone_prb}. The interplay
of electron-electron $(e-e)$ and electron-lattice $(e-l)$ interaction, in particular the coupling with the buckling mode associated to
rotation of the oxygen cages ubiquitous in perovskyte oxides, causes an effective
dimerization via an orbital striction mechanism, also responsible for the predicted OO
pattern\cite{gianluca-lcmo,colizzi}.
Quite surprisingly, OO combined with the forementioned effective dimerization
appears to be the main source of an electronically induced polarization, since
CO is strongly inihibited in the dimerized state\cite{barone_prb}.

In our opinion, these findings call for further research and suggest that the role of OO in establishing or contributing to
ferroelectric polarization deserves a deeper understanding.
In order to provide a clear picture of the possible general mechanisms by which $\bm P$
can be induced by OO, therefore, we analyze  within
the BP approach a simple one-dimensional tight-binding model where OO can be controlled by a Hubbard-like interaction.
We will consider only spinless fermions, which allow to easily disentangle the role of spin and
charge from orbital degrees of freedom in causing a finite polarization.
We will show that bond inequivalency without CO but with a finite orbital-occupancy
disproportionation is enough to lead to a ferroelectric state, provided the intra-orbital hopping interactions are
different. 
The choice of this extremely simple model is motivated by the clear insight it provides within the BP
framework, rather than by its straight application to some real material. Nonetheless, we expect the proposed mechanism
to be rather general and possibly relevant for the recently proposed multiferroic vanadium spinel
CdV$_2$O$_4$\cite{giovannetti2011} and similar systems, as we discuss in our concluding remarks.

{\it Model and analytical formulas.} 
The possible interplay between orbital degrees of freedom and bond inequivalency in inducing a bulk $\bm P$ can be
addressed by considering a spinless two-band Peierls-Hubbard model, defined on a one-dimensional chain, whose
Hamiltonian reads:
\beq
H &=& \sum_{j,\gamma,\gamma'}\,t_{j}^{\gamma\gamma'}\,(d^\dagger_{j\gamma}
d^{\phantom{\dagger}}_{j+1\gamma'}+h.c.)\,
+\frac{K}{2}\,\sum_j\,u_j^2 \nonumber\\
&+& U\sum_j\, d^\dagger_{j1}d^{\phantom{\dagger}}_{j1}
d^\dagger_{j2}d^{\phantom{\dagger}}_{j2},
\eeq
with $d^\dagger_{j\gamma}$ creating an electron in the orbital $\gamma$ of site $j$.
The hopping parameters are modulated by the $e-l$ interaction, with coupling constant $\alpha$,
as $t_{j}^{\gamma\gamma'}= t^{\gamma\gamma'}\left[1+\alpha(u_j-u_{j+1})\right]\,$, where $u_j$ are the
displacement of the ions from their equilibrium position along the chain, with the associated elastic constant $K$. Without
loss of generality, in the
following we will neglect any inter-orbital hopping, assuming $t^{\gamma\gamma'}=\delta_{\gamma,\gamma'}t_\gamma$ with
$t_1\neq t_2$. The
last term describes the Coulomb interaction between electrons occupying different orbitals, parametrized by the Hubbard
constant $U$.

Let us consider the half-filling case with average local occupation fixed to $n=1$\cite{footnote1}. 
Peierls-like dimerization can be realized via a staggered pattern of distortions $u_j=(-1)^j\,u_0$, with $u_0$ to be
variationally determined, whereas the Coulomb interaction can be decoupled via standard Hartree-Fock linearization. The
model is then formally equivalent to a single-band spinful Peierls-Hubbard model, whose staggered magnetization
$m=(-1)^j (n_{j\uparrow}-n_{j\downarrow})$ maps to the difference between average orbital occupancies
$\Delta= (-1)^j(n_{j1}-n_{j2})$, that may serve
as the order parameter for orbital ordering. Since the
periodicity of dimerization and of staggered OO is the same, the Hamiltonian in reciprocal space can be defined in the
reduced Brillouin zone, reading
\beq\label{hamk}
h_k^\gamma=\left(\begin{array}{cc}
\varepsilon_\gamma(k) +\frac{1}{2}U& \delta_\gamma(k)+(-1)^\gamma\frac{1}{2}U\Delta\\
\delta_\gamma^*(k)+(-1)^\gamma\frac{1}{2}U\Delta & -\varepsilon_\gamma(k) +\frac{1}{2}U\\
\end{array}
\right)
\eeq
in each orbital sector, where $\varepsilon_\gamma(k)=-2t_\gamma\cos k$ and $\delta_\gamma(k)=i 4 \alpha\, u_0\,  t_\gamma\sin k$,
with both $u_0, \Delta$ to be self-consistently determined. Eigenstates are given by:
\beq\label{eq:eigennnvalues}
E^\gamma_\pm(k)=\frac{1}{2}U\pm\sqrt{\varepsilon_\gamma(k)^2+\vert \delta_\gamma(k)\vert^2+\frac{1}{4}\left(U\Delta\right)^2},
\eeq
implying an insulating groundstate for nonvanishing $u_0$ or $\Delta$, which at half filling must fulfill the following self-consistency
equations:
\beq
\frac{4\pi}{U} &=&\sum_{\gamma,k}\left[\varepsilon_\gamma(k)^2+\vert
\delta_\gamma(k)\vert^2+\frac{1}{4}(U\Delta)^2\right]^{-1/2},\\
\frac{\pi K}{8\alpha^2} &=&\sum_{\gamma,k}\frac{t_\gamma^2\sin^2k}{\sqrt{\varepsilon_\gamma(k)^2+\vert
\delta_\gamma(k)\vert^2+\frac{1}{4}(U\Delta)^2}}.
\eeq 

\begin{figure}[b]
\includegraphics[width=\linewidth]{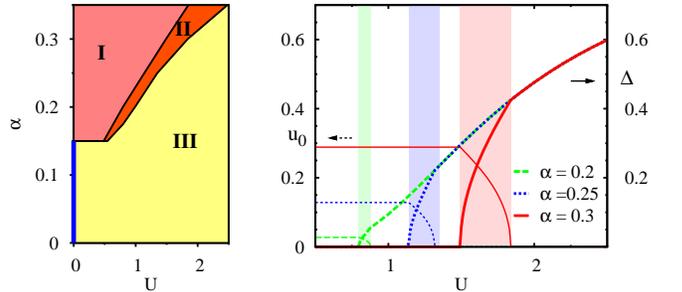}
\caption{Left: Phase diagram in the $U-\alpha$ space for the two-band spinless Peierls-Hubbard with inequivalent hoppings
$t_1/t_2=3$: the thick line represents the metallic solution found at $U=0$, I and III label respectively the Peierls and
OO insulator, with II the mixed-character insulator intruding between them. All parameters are expressed in units of
$t=(t_1+t_2)/2=1$ and the elastic energy is fixed to $K=1$. Right: evolution of the
dimerization parameter $u_0$ (thin lines) and of the orbital-occupancy disproportionation $\Delta$ (thick lines)
as a function of $U$ for different values of the $e-l$ interaction.
Regions where OO coexists with structural distortion are highlighted.}\label{fig1}
\end{figure}

Depending on the coupling constants $\alpha, U$, four phases are found characterized by different structural and orbital
properties. For small $\alpha$ and $U=0$, we found a metallic ground-state, with no orbital-occupancy disproportionation and no
structural distortion. As $U,\alpha$ are increased, three different insulating ground-states develop with finite $\Delta$
or $u_0$. A first insulating  phase, with $\Delta\neq0$ and $u_0=0$, is induced by increasing the $e-e$ interaction,
exactly mapping
onto the antiferromagnetic insulator found for the half-filled single-band Hubbard model. On the other hand, a
dimerized state, with $u_0\neq0$ and $\Delta=0$, develops with increasing $\alpha$ at small $U$, which is the
conventional Peierls insulator with structural distortions. A third insulating phase
intrudes between the first two at moderate values of $U$ by increasing the $e-l$ interaction;
this insulating state is
characterized by the coexistence of OO and structural dimerization. The phase diagram in the $U-\alpha$ space
is shown in Fig. \ref{fig1} for a given ratio between hopping parameters, namely $t_1/t_2=3$. The evolution of
dimerization $u_0$ and orbital occupancy disproportionation $\Delta$ is also shown as a function of $U$, highlighting the
coexistence regions at different values of the $e-l$ interaction. Even if the phase boundaries depend on the
ratio $t_1/t_2$  and may be affected by the mean-field approximation,
the phase diagram is qualitative the same as the one obtained for the single-band Peierls-Hubbard
model by more accurate numerical methods\cite{sanvito2010}. 

\begin{figure}
\includegraphics[width=\linewidth]{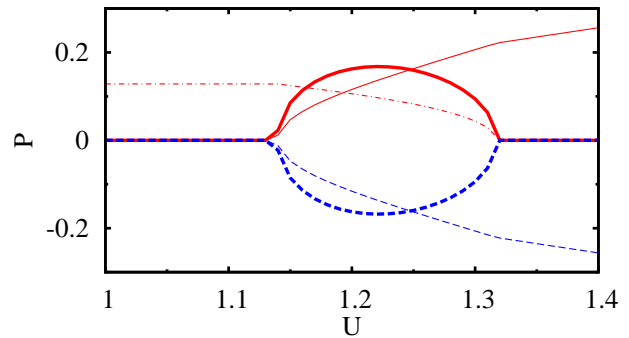}
\caption{Evolution of the Berry-phase polarization (thick lines) as a function of $U$ at $\alpha=0.25$ and $t_1/t_2=3$. 
Thin lines are the dimerization parameter $u_0$ (dot-dashed line) and the OO parameter $\Delta$ (solid or dashed);
dashed lines correspond to the swapping of orbital occupancy by keeping fixed
the structural distortion of the chain, causing a reversal of $P$.}\label{fig2}
\end{figure}

Turning to ferroelectricity, the key quantity to be considered is the overlap matrix defined as $
S_{m,m^{\prime}}(k,k^{\prime})=\langle \psi_{m,k}\vert e^{-i\frac{2\pi}{L}\hat{x
}}\vert \psi_{m^{\prime},k^{\prime}}\rangle
$, which allows to evaluate the expectation value of the position operator $\hat{x}$ on the eigenstates
$\psi_{m,k}$, with $m$ the occupied band index, which obey periodic boundary conditions on a $1d$ chain of lenght $L$\cite{resta98}.
Since the overlap matrix has nonvanishing elements only when each pairs of $k,k^{\prime}$ differs by $\epsilon=2\pi/L$, one
can explicitly evaluate its elements on the eigenstates of Hamiltonian (\ref{hamk}) in the limit of $L\rightarrow\infty$, finding
$S_{m,m^{\prime}}(k,k+\epsilon)=\delta_{m,m^{\prime}}(1+i\epsilon\,S_m)$ with:
\beq\label{eq:smatrix}
S_m=(-1)^m\,\frac{4\,\varepsilon_m(k)^2}{E^m_+-E^m_-}\,\frac{\alpha\,u_0\,U\,\Delta}{(U\Delta)^2+4\vert
\delta_m(k)\vert^2}
\eeq
where $m=1,2$ labels the occupied bands with eigenvalues $E^1_-(k),E^2_-(k)$ as given in Eq. (\ref{eq:eigennnvalues}). 
By assuming an uniformly charged ionic background, no ionic contribution arises from the chain distortion in the absence of
charge disproportionation; polarization, then, has only an electronic origin and
is evaluated as\cite{resta98}:
\beq\label{eq:panalytic}
\frac{\pi}{e}P = -\mbox{Im}\,\sum_k\,\log\det S(k,k+\epsilon)\simeq-\int_{-\pi/2}^{\pi/2}\sum_m S_m\,dk,\nonumber\\
\eeq
$e$ being the electron charge.
By inspecting Eqs. (\ref{eq:smatrix}),(\ref{eq:panalytic}) it comes clear that two conditions must be simultaneously 
fulfilled in order to have a nonvanishing $P$: i) both OO  and structural dimerization, with $u_0\Delta\neq 0$, must occur, and ii) hopping
integrals must be different in different orbital channels, i.e. $t_1\neq t_2$, otherwise the two addends in the r.h.s of
Eq. (\ref{eq:panalytic}) would cancel out exactly leading to $ P=0$. The explicit expression for $S_m(k)$, as given in Eq.
(\ref{eq:smatrix}), also implies that
the polarization can be reversed either by swapping the orbital occupancy between the two inequivalent sites in the doubled
cell or by interchanging short and long bonds of the distorted chain. A paradigmatic evolution of $P$ in the dimerized
orbital-ordered state is shown in Fig. (\ref{fig2}). Enforcing the correspondence between the two-band spinless and one-band spinful
Peierls-Hubbard model, one immediately recognizes that the second condition is never fulfilled in the spinful model, since
hopping integrals are spin independent and identical. This explains why the antiferromagnetic ground state of the single-band
Peierls-Hubbard model is never ferroelectric unless a CO is
realized, e.g., by including a staggered potential, as already discussed in Ref. \onlinecite{gianluca-ttfca}.

\begin{figure}
\includegraphics[width=\linewidth]{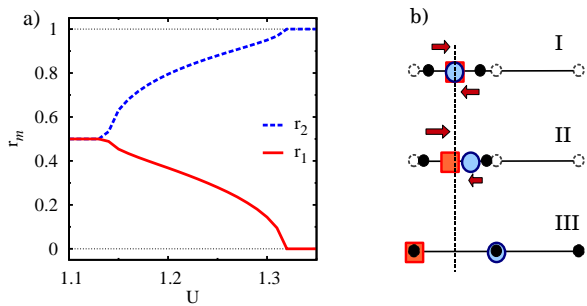}
\caption{$a)$ Evolution of the Wannier function centers $r_m$, in
units of lattice spacing $a_0=1$, as a function of $U$ for $\alpha=0.25$ and $t_1/t_2=3$. b)
Schematic picture of the displacement of WF centers as the system evolves from the Peierls insulator (I) to
the OO insulator (III) through the dimerized orbital-ordered state (II), highlighting the inequivalent local dipoles that
develops in phase II (empty circles indicate the undistorted position of sites along the chain; the displacements are exaggerated to make the picture clearer).}\label{fig3}
\end{figure}

A pictorial interpretation of the origin of the electronic $\bm P$ in the orbital-ordered dimerized state
is provided by looking at the center of the
Wannier functions (WF) of the two occupied bands, that are easily evaluated for the $1d$ system under
consideration. As pointed out in Ref.
\onlinecite{marzari}, in fact, the
construction of maximally localized WFs is not needed in the present case; WF centers can be obtained
as the eigenvalues of a matrix $\Lambda$, constructed as the product
of the unitary parts of the $S$ matrices along the k-point
string (given by the matrix product $VW^\dagger$
taken from the singular value decomposition $S=V\Sigma W^\dagger$,
where $V$ and $W$ are unitary and $\Sigma$ is a diagonal matrix with
nonnegative diagonal elements). In the limit of infinite chain (with $\epsilon\rightarrow 0$), it turns out that the WF
centers $r_m$ are given by the two addends in the r.h.s of Eq. (\ref{eq:panalytic}), a well-known result in the context of
the modern theory of polarization\cite{berryp_rev}. That means that in the orbital-ordered dimerized state
the two WF centers move in opposite directions along the chain, reflecting the staggered orbital occupancy; the different
hopping integrals $t_1\neq t_2$ (implying $\vert S_1\vert\neq\vert S_2\vert$), however, make the displacements of $r_m$ with respect to their
nonpolar positions inequivalent, giving rise to uncompensated local dipoles that sum up resulting in a bulk $P$. This is shown in Fig.
\ref{fig3}, where the evolution of $r_m$ is plotted as a function of $U$ for $\alpha=0.25$ and $t_1/t_2=3$. In
the Peierls insulator, where bonds differentiate in alternating short and long ones,
both the WF centers lie in the middle of the short bond. On the other hand, when OO develops, e.g. with prevalent
$\gamma=1$ character on odd sites and $\gamma=2$ on even sites, the corresponding WF centers move towards the most
occupied sites with same orbital character. Due to the different values of hopping integrals, these displacements are
inequivalent and cause the local dipoles to be uncompensated. Eventually, when the structural distortion disappears in the
OO insulator,
each $r_m$ lies on a site of the chain, as schematically shown in Fig. \ref{fig3}b).

{\it Conclusions.} We analysed the ferroelectric properties of a simplified model in the framework of the modern theory of
polarization in order to unveil the necessary conditions by which OO phenomena may lead to an electronic
$ bm P$. The spinless two-band Peierls-Hubbard model allowed us to consider the cooperative interplay between
OO and Peierls-like dimerization along a $1d$ chain. The derived analytical formulas, even if obtained at a
mean-field level, clarify that
$P$ may arise from OO only if a concomitant bond inequivalency (appearing as a structural
dimerization in the considered model) develops. Na\"ively, the emerging $P$ may be interpreted as a superposition of two
inequivalent charge-localization phenomena in different orbital sectors. By looking at Wannier-function centers, that can
be viewed as the localization centers of the continuous electronic charge distribution, local electric dipoles
develop as the WF centers, with given orbital character, move towards the site with most orbital character of the same
kind; in the presence of bond inequivalency, these dipoles are uncompensated, giving rise to a nonzero $P$. Finally, the
polarization can be reversed by reversing OO or the chain dimerization. The unveiled
mechanism is different from the one previously reported for undoped manganites\cite{barone_prl}.
In that case, Jahn-Teller lattice distortions tuned OO, whereas a large Hund coupling in the E-type antiferromagnetic
phase constrained electrons to hop along ferromagnetic zig-zag chains; the change of electron motion around each site
displaying OO was responsible, through direction-dependent {\it inter-orbital} hopping processes, for the phase change of
the Bloch wavefunctions leading to $P$.

Even though the analysis has so
far been restricted to little more than a toy model, the proposed mechanism
may prove relevant for spinel crystals, such
as MgTi$_2$O$_4$\cite{schmidt}, CuIr$_2$S$_4$\cite{radaelli} or vanadium spinels as ZnV$_2$O$_4$\cite{pardo} or
CdV$_2$O$_4$\cite{giovannetti2011}. In the simplest approximation, the electronic structure of these systems 
has essentialy one-dimensional character,
stemming from the strong direct $d-d$ overlap of the $B-$site $t_{2g}$ orbitals. The strong
anisotropy of the $d-$electron distribution, in fact, is such that electrons in $xy, yz, xz$ orbitals would hop preferably
in spinels along $xy,yz,xz$ directions respectively, implying also strongly direction-dependent (orbital-dependent) hopping
integrals along each effective one-dimensional ``chain''\cite{khomskii2011}. For instance, in
vanadates (V$^{3+}$) one finds that the $xy-$orbitals are
always occupied due to a tetragonal distortion while the remaining $t_{2g}$ electron moves in degenerate $yz,zx$ bands;
adopting Slater-Koster parametrization, the hopping amplitudes along a single one-dimensional ``chain'', e.g., parallel
to $yz$ direction would fulfill $t_{yz,zx}=0$ and $t_{yz,yz}\simeq 2.8t_{zx,zx}$\cite{SK,harrison}, close to our choice
of parameters.
Each chain in the spinel structure
may then be subject to Peierls transition, which has been suggested to be strongly tied to possible orbital ordering
phenomena\cite{khomskiimizokawa2005, saha-dasgupta2010}, eventually driving the system insulator. Interestingly, at least
one of the forementioned vanadium spinels, CdV$_2$O$_4$, has been reported to be ferroelectric\cite{giovannetti2011}. Other
issues should be taken into account if willing to extend our analysis to more realistic situations, such as the
forementioned spinels. In particular, the role of spin degrees of freedom, that has been overlooked in order to highlight
the role of orbital degrees of freedom, may play a relevant role in inducing the bond inequivalency, analogously to what
happens in Ca$_3$CoMnO$_6$ or in collinear rare-earth manganites $R$MnO$_3$. Furthermore, the interplay
between spin and orbital degrees of freedom could be relevant for novel multiferroic or magnetoelectric effects.

This work has been supported by the European Community's Seventh Framework Programme FP7/2007-2013 under grant agreement
No. 203523-BISMUTH.


\begin{thebibliography}{99}
\bibitem{footnote1}Even though this choice may not lead to the real ground state as long as $t_1/t_2\neq 1$, it allows
to disentangle the role of orbital and charge ordering as the source for bulk polarization.
\end{thebibliography}
\end{document}